\documentclass[showpacs,amsmath,amssymb,twocolumn,floatfix]{revtex4}

\usepackage{graphicx}
\usepackage{amssymb}
\usepackage{amsmath}

\newcommand{\ket}[1]{|#1\rangle}

\begin{document}

\title{Intensity fluctuations in steady state superradiance}

\author{D. Meiser and M. J. Holland}
\affiliation{JILA and Department
  of Physics, The University of Colorado, Boulder, Colorado
  80309-0440, USA}

\date{\today}

\begin{abstract}
  Alkaline-earth like atoms with ultra-narrow optical transitions
  enable superradiance in steady state.  The emitted light promises to
  have an unprecedented stability with a linewidth as narrow as a few
  millihertz.  In order to evaluate the potential usefulness of this
  light source as an ultrastable oscillator in clock and precision
  metrology applications it is crucial to understand the noise
  properties of this device.  In this paper we present a detailed
  analysis of the intensity fluctuations by means of Monte-Carlo
  simulations and semi-classical approximations.  We find that the
  light exhibits bunching below threshold, is to a good approximation
  coherent in the superradiant regime, and is chaotic above the second
  threshold.
\end{abstract}


\maketitle

\section{Introduction}
\label{sec:intro}

Dicke superradiance is a paradigmatic collective effect in cavity
quantum electrodynamics (QED).  At the fundamental, level
superradiance is a quantum interference effect in which the
probability amplitudes for the emission of a photon from several atoms
conspire to yield a collective light emission rate that is larger than
for uncorrelated atoms.  Due to its great conceptual simplicity and
generality superradiance has been extensively studied both
experimentally and theoretically.  The noise properties of the light
emitted in superradiance have been of particular interest.  This is
because the early stages of superradiance are often initiated by
quantum fluctuations which are subsequently amplified by the
collective emission
process~\cite{haake79,Carmichael:StochasticInitiation}.  Superradiance
can thus serve as a physical phenomenon that allows us to study the
microscopic quantum fluctuations through their macroscopic
consequences.  Examples of this macroscopic manifestation of the
quantum fluctuations are the first passage time statistics of the
superradiant pulse~\cite{haake81a, Haake:DelayTimeStatisticsPRL,
  Hermann:passagetimestatistics} as well as the second order
correlations of the field~\cite{Bonifacio:QuantumStatTheorySR1}.

Nearly all realizations of Dicke superradiance have been in the pulsed
regime.  In contrast to such experiments we have recently proposed a
system based on earth-alkaline-like atoms in which superradiance can
be achieved in steady state
\cite{Meiser:SrLaser,Meiser:SteadyStateSuperradiance}.  The interest
in that light source derives from the extremely narrow linewidth of
the generated light.  For experimentally realizable parameters
linewidths in the millihertz range could potentially be realized.
The light generated this way could thus serve as an ultra-stable local
oscillator with a stability that is about two orders of magnitude
better than the current state-of-the-art.  At the core of this device
are atoms with an ultra-narrow optical transition coupled to a high
finesse cavity.  The atoms collectively emit photons into the cavity
mode and they are concurrently repumped to the excited state,
providing a steady supply of energy.  The collective decay of the
atoms via the cavity mode establishes a collective atomic dipole,
which radiates much more strongly than independent atoms would.
Depending on the repumping rate, the system can also exhibit
subradiance, or thermal light as would be the case for an
ensemble of random radiators. Qualitatively similar behavior has been
predicted for the overdamped many-atom micromaser
\cite{Temnov:OverdampedMicromaser}.

Just as for pulsed Dicke superradiance, the higher order correlations
of the light are non-trivial and a solid knowledge of them is crucial
for a full understanding of the collective light generation mechanism
as well as for potential applications.  Correlations of the intensity
can be used to quantitatively study these fluctutations and these are
the subject of this paper.  Some aspects of the noise properties of
continuously pumped, collectively emitting systems have also been
discussed recently in the context of collectively radiating low
dimensional solid state
systems~\cite{Temnov:SuperradianceAndSubradiance,
  Scheibner:SuperradianceQuantumDots}.  For instance Temnov and Woggon
studied the photon statistics deep in the subradiant regime
in~\cite{Temnov:PhotonStatistics}.

The goal of this paper is to fully characterize the intensity
correlations of the light generated by means of steady state
superradiance.  To this end we introduce a simplified model that
captures all the essential aspects of the problem in section
\ref{sec:model} and we recall the basics of steady state
superradiance.  Section \ref{sec:results} presents our results on the
Hanbury Brown-Twiss correlations of the generated light obtained for
small atom numbers using quantum Monte-Carlo simulations and
semiclassical approximations in the limit of large atom numbers.

\section{Model}
\label{sec:model}

Calculating higher order correlations of quantum fields is typically a
hard problem.  Analytic closed form solutions are known in only a
few special cases and thus we have to rely mostly on numerical
simulations.  Because of the exponential scaling of the dimension of
the Hilbert space of the system with the number of atoms we must
restrict ourselves to the simplest model that still captures the
essential physics that we are interested in.

\begin{figure}
  \includegraphics{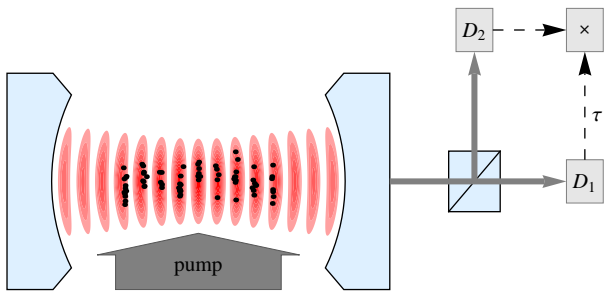}
  \caption{(Color online) Schematic illustration of $N$ two level
    atoms confined in a single mode cavity field indicated by the
    black dots.  The atoms are being incoherently repumped.  The
    output of the cavity field is monitored by two detectors $D_1$ and
    $D_2$ with a variable time delay $\tau$ between them.}
  \label{fig:schematic}
\end{figure}

The core ingredients of a system exhibiting steady state superradiance
are illustrated in Fig. \ref{fig:schematic}.  An ensemble of $N$ two
level atoms with excited state $\ket{e}$, ground state $\ket{g}$, and
transition frequency $\omega_a$ are collectively coupled to a high
finesse cavity with resonance frequency $\omega_c$.  The atoms are
independently repumped from the ground state to the excited state in
order for the atoms to be able to radiate photons continuously.

The non-collective nature of the repumping is essential for two
reasons.  First, it is much easier to achieve experimentally than
collective repumping, for instance by optical pumping through an
auxiliary excited state.  Second, and somewhat paradoxically, it can
balance the effects of other incoherent processes such as spontaneous
emission and dephasing that would normally drive the atoms into less
collective states that do not exhibit superradiance.  A purely
collective repumping as considered
e.g. in~\cite{Haake:SuperradiantLaser,
  Haake:NoiseReductionStationarySuperradiance} cannot change the
length of the atomic Bloch vector.  Therefore, if the length decays
due to dissipative processes, it cannot be restored and the
superradiance stops.  In contrast, the length of the atomic Bloch
vector can grow in the case of non-collective pumping so that a
collective atomic dipole can develop from an ensemble of completely
independent atoms.

\subsection{Atom-field master equation}
\label{subsec:AtomFieldMasterEquation}

Mathematically, the coupled atom cavity system can be described by the
Hamiltonian
\begin{equation}
  \hat H = \hbar \omega_a \hat J_z + \hbar \omega_c \hat a^\dagger \hat a
  + \frac{\hbar g}{2}(\hat a^\dagger \hat J_- + \hat J_+ \hat a)\;.
  \label{eqn:AtomCavityHamiltonian}
\end{equation}

Here we have introduced an angular momentum representation for the
atoms in the usual way by identifying the excited state with the spin
up state of a fictitious spin 1/2 system and the ground state with the
spin down state.  The operators
\begin{equation}
\hat J_z = \frac{1}{2}\sum_{j=1}^N \hat \sigma_z^{(j)}
\end{equation}
and
\begin{equation}
\hat J_- = \hat J_+^\dagger = \sum_{j=1}^N\hat \sigma_-^{(j)}
\end{equation}
are the $z$-component and ladder operators of the total angular
momentum.  In these equations $\hat \sigma_z^{(j)}=\ket{e}\langle
e|-\ket{g}\langle g|$ is a Pauli matrix pertaining to atom number $j$
and $\hat \sigma_-=\hat \sigma_+^\dagger=\ket{g}\langle e|$ is the
spin-flip operator for atom $j$.  The operators $\hat a$ and $\hat
a^\dagger$ are the annihilation and creation operators for a photon in
the cavity mode.  For simplicity we assume that the coupling constant
$g$ of the atoms to the cavity is identical for all atoms.  This could,
in principle, be achieved by trapping the atoms at the antinodes of the
cavity mode.  Less ideal spatial configurations of the atoms merely
lead to a reduction of the effective number of atoms by a factor of
order one that clearly has no impact on the basic conclusions of this
paper.

Besides the coherent interaction of the atoms with the cavity field
there are also dissipative processes due to the coupling of the atoms
to field modes outside of the cavity and due to decay of the
cavity fields.  The decay of the cavity can be accounted for with the
usual Born-Markov master equation for the reduced density matrix for
atoms and cavity field,
\begin{equation}
  \frac{d\hat \rho}{dt}=
  \frac{1}{i\hbar}[\hat \rho,\hat H] +
  \mathcal{L}_{\rm cav}[\hat \rho] +
  \mathcal{L}_{\rm pump}[\hat \rho]\;,
  \label{eqn:MasterEqn}
\end{equation}
where the Liouvillian for the cavity decay with intensity decay rate
$\kappa$ is
\begin{equation}
  \mathcal{L}_{\rm cav}[\hat \rho] = 
  -\frac{\kappa}{2}
  (\hat a^\dagger \hat a \hat \rho +
  \hat \rho \hat a^\dagger \hat a -
  2 \hat a \rho \hat a^\dagger)\;.
  \label{eqn:Liouvillian}
\end{equation}
The repumping of the atoms with pump rate $w$ is described by the
Liouvillian
\begin{equation}
  \mathcal{L}_{\rm pump} [\hat\rho]=
  -\frac{w}{2}\sum_{j=1}^N(
  \hat \sigma_-^{(j)}\hat \sigma_+^{(j)}\hat \rho +
  \hat \rho \hat \sigma_-^{(j)}\hat \sigma_+^{(j)} -
  2 \hat \sigma_+^{(j)}\hat \rho \hat \sigma_-^{(j)}) \;.
\end{equation}
We are assuming that the spontaneous emission of the atoms into free
space with decay rate $\gamma$ can be neglected.  In general this
assumption requires $\mathcal{C}\gg 1$ where $\mathcal{C} =
g^2/(\kappa\gamma)$ is the single atom cooperativity parameter, see
Eq. (\ref{eqn:gammac}) below.  Note however that in the superradiant
regime where the decay through the cavity is collectively enhanced it
is found that the much less stringent condition $N \mathcal{C}\gg 1$
is sufficient for this approximation to be justified.

In order to calculate correlation functions of the generated light
field we simulate the dynamics of the system subject to the master
equation (\ref{eqn:MasterEqn}) using the Monte-Carlo wavefunction
technique
\cite{Dalibard:QuantumJumpApproach,
Zoller:QuantumJumpApproach,plenio1998qja}.
In that technique the evolution of the system is represented by an
ensemble of stochastic wavefunction trajectories $\{\ket{\psi(t)}\}$
where each trajectory $\ket{\psi(t)}$ is a representative evolution of
the system.

\subsection{Adiabatic elimination of field in ``bad cavity'' limit}
\label{SubSec:AdiabaticElimination}

The Hamiltonian Eq. (\ref{eqn:AtomCavityHamiltonian}) is suitable for
the Monte-Carlo simulations because it directly grants access to the
field correlations we are interested in.  For analytical calculations
it is desirable to further simplify the problem by exploiting the fact
that the cavity field decays so much faster than the atomic coherence.
Adiabatically eliminating the light field yields the effective
superradiance master equation~\cite{Bonifacio:QuantumStatTheorySR1},
\begin{align}
    \frac{d\hat \rho}{dt}&=
    -\frac{\Gamma_c}{2}\left( \hat J_+\hat J_-\hat
      \rho+\hat \rho \hat J_+\hat J_- -2 \hat J_- \hat \rho \hat
    J_+ \right) \label{eqn:masterEqnJustAtoms} \\
  &\quad -\frac{w}{2}\sum_{j=1}^N\left(\hat \sigma_-^{(j)}\hat
    \sigma_+^{(j)} \hat \rho + \hat \rho\hat
    \sigma_-^{(j)}\hat \sigma_+^{(j)}-2 \hat \sigma_+^{(j)}
    \hat \rho  \hat \sigma_k-^{(j)}\right)\;.\notag
\end{align}
Here, the collective decay rate of the atoms is given by
\begin{equation}
  \Gamma_c = \mathcal{C} \gamma = g^2/\kappa\;.
  \label{eqn:gammac}
\end{equation}
The condition for the validity of the adiabatic elimination of the
cavity field is that the cavity field relaxes much faster as the
atoms.  Using that the fastest atomic relaxation rates are obtained in
the superradiant regime and that they are of order
$N\mathcal{C}\gamma$ we find the quantitative condition
\begin{equation}
\kappa \gg N\mathcal{C}\gamma\;.
\end{equation}
In this ``bad cavity'' limit the role of the cavity mode reduces to
providing a collective decay channel for the atoms.  The
simplification brought about by the elimination of the field is
two-fold.  It allows us to deal with the atomic degrees of freedom
only and all parameters of the full coupled atom-cavity system have
collapsed into just one characteristic parameter, $w/\Gamma_c$.

\subsection{Hanbury-Brown-Twiss signal}

Fluctuations of the light intensity can be characterized
experimentally by a Hanbury-Brown-Twiss-like setup as illustrated in
Fig. \ref{fig:schematic} \cite{Glauber:Optical_Coherence1,mandel95}.
The light passes through a 50/50 beam splitter and the intensities in
each output-port of the beam-splitter is detected with a photo-diode.
The photo-currents of each detector are then multiplied and
integrated.  A variable delay $\tau$ can be imparted on one of the
outputs of the beam splitter in order to measure correlations of the
field at different times.  Using such a setup it is possible to
measure the joint probability $P_2(t, t+\tau)\Delta t \Delta\tau$ to
detect a photon both in a time interval $\Delta t$ at $t$ and in a
time interval $\Delta \tau$ at $t+\tau$.  According to the theory of
photo-detection \cite{Glauber:Optical_Coherence2,mandel95} this
probability can be calculated in terms of normally ordered expectation
values of the field amplitude,
\begin{align}
  g^{(2)}(t,\tau)&\equiv \notag
  \frac{P_2(t, t+\tau)\Delta t \Delta\tau}{
    P_1(t)\Delta t P_1(t+\tau)\Delta\tau}\\
&=\frac{\langle \hat a^\dagger(t)\hat a^\dagger(t+\tau)
    \hat a (t+\tau)\hat a\rangle
  }
  {\langle \hat a^\dagger(t)\hat a(t)\rangle
    \langle\hat a^\dagger(t+\tau)\hat a(t+\tau)\rangle
  }\;.
  \label{eqn:g2Definition}
\end{align}
We have normalized the joint probability to the single time
probabilities for photon detection, $P_1(t)\Delta t$ and
$P_1(t+\Delta\tau)\Delta \tau$.  In steady state, which is the case of
primary interest here, $g^{(2)}(t,\tau)$ does not depend on $t$.  For
notational convenience we drop the variable $t$ from
$g^{(2)}(t,\tau)$, i.e. we simply write $g^{(2)}(\tau)$.  In writing
Eq.~(\ref{eqn:g2Definition}) we have also made use of the result from
the input-output theory for cavities that normally ordered correlation
functions outside of a cavity are equal to the normally ordered
correlation functions of the intra-cavity field, provided that the
input ports of the cavity are in vacuum
states~\cite{Gardinger:QuantumNoise,Walls:QuantumOptics}.

The second order correlation function at zero time delay,
$g^{(2)}(0)$, is related to the fluctuations $\Delta I^2 = \langle
\hat I^2 \rangle -\langle \hat I\rangle^2$ of the out-coupled photon
flux $\hat I$.  These fluctuations $\Delta I^2$ can be used to
characterize the instability of the intensity of the out-coupled beam
because $\hat I$ is proportional to the beam intensity.  Typical photo
detectors have a detection bandwidth $B$ that is extremely large
compared to the cavity bandwidth and in that case the fluctuations of
the intensity are given by
\begin{equation}
  \Delta I^2 = I^2(g^{(2)}(0)-1) + B I\;,
\end{equation}
where $I=\langle \hat I\rangle$.  For coherent light the arrival of
photons at the detectors are a Poisson process in which the arrival
times are completely random with mean rate of arrival $I$. For such
light $\Delta I^2 = B I$ and we have $g^{(2)}(0)=1$.  Light with
larger intensity fluctuations is called super-Poissonian and light
with smaller intensity fluctuations is called sub-Poissonian.
Super-Poissonian light has $g^{(2)}(0)>1$ and photons arrive in
bunches, while sub-Poissonian light has $g^{(2)}<1$ and photons are
anti-bunched, i.e. they arrive more regularly than pedicted by Poisson
statistics.

Our numerical wave-function Monte-Carlo simulations grant us access to
correlation functions in two ways.  First, we can calculate
expectation values of \emph{system} observables $\hat O$ in the usual
way by calculating $\langle \psi |\hat O \ket{\psi}$ and averaging
over the ensemble of trajectories.  Alternatively we can extract the
correlation functions by an analysis of the decay times of the system
that very closely mirrors an actual experimental procedure.  This
latter method is easier to implement for non-zero time delays.  All
results on the intensity correlations presented here were calculated
this way due to the greater flexibility of having access to zero
and non-zero delays.
\begin{figure}
  \includegraphics{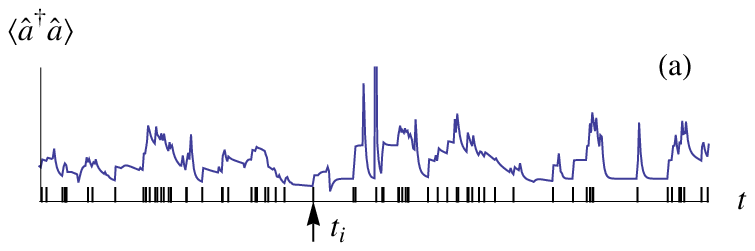}\\
  \includegraphics{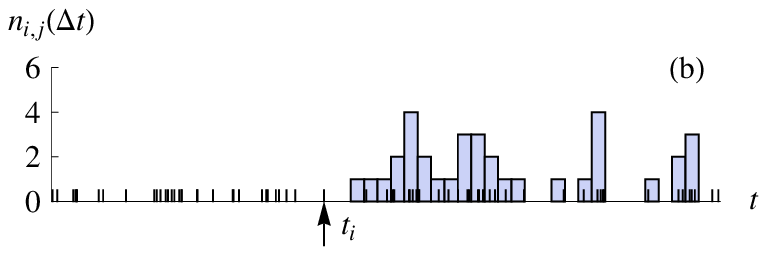}
  \caption{(Color online) Time evolution of the intracavity intensity
    during a single Monte-Carlo trajectory (a) for $N=10$ atoms and
    $w=5 \Gamma_c$. The black ticks indicate cavity decay
    events. Panel (b) shows the binning of the decay times subsequent
    to $t_i$ leading to the histogram $n_{i,j}(\Delta t)$ used in the
    calculation of $g^{(2)}(\tau)$ (see text for explanation).}
  \label{fig:schematicG2Calculation}
\end{figure}

The procedure by which we calculate $g^{(2)}(\tau)$ from the cavity
decay times is illustrated in Fig. \ref{fig:schematicG2Calculation}.
Part a) of that figure shows the evolution of the mean photon number
inside the cavity for an example trajectory.  Cavity decay events are
indicated by the black ticks at times $t_i$.  For each photon emission
event we calculate the Histogram $n_{i,j}(\Delta t) =
\#\text{photons in }(t_i+j\Delta t,t_i,+(j+1)\Delta t]$.  By averaging
these histograms over all $i$ we find $\bar{n}_j(\Delta t)=n_{\rm
  phot}^{-1}\sum_i n_{i,j}(\Delta t)$, where $n_{\rm phot}$ is the
total number of photons emitted.  This histogram is closely related to the
conditional probability $P(t+j\Delta t|t)$ to find a second photon at
time $t+j\Delta t$ provided that a first photon has been detected at
time $t$, {\it i.e.} $\bar{n}_j(\Delta t) = \Delta t P(t+j\Delta
t|t)$.  Using the relation between that conditional probability and
the joint probability, $P_2(t,t+j\Delta t)=P(t+j\Delta t|t)P_1(t)$, we
can then find $g^{(2)}(j\Delta t)$ on a grid with spacing $\Delta t$
according to
\begin{equation}
  g^{(2)}(j\Delta t)=
  \frac{\bar{n}_j(\Delta t)}{n_{\rm phot}/\#\text{bins}}\;,
\end{equation}
where $n_{\rm phot}/\#\text{bins}=P_1(t)\Delta t$ is the mean number
of photons per bin.  The choice of the bin width $\Delta t$ is a trade
off between resolution and statistical fluctuations and it has to be
chosen differently for different simulation parameters.  It must be
small enough to resolve the dynamics of the system.  Once that
constraint is satisfied it should be as long as possible in order to
yield the smallest fluctuations.

\section{Results}
\label{sec:results}

Generally, as explained in detail in
\cite{Meiser:SteadyStateSuperradiance}, three regimes of light
emission can be distinguished depending on the repump strength.  If
the repump rate is smaller than the effective atomic decay rate, $w\ll
\Gamma_c$, the atoms evolve into a dark state in which the emission of
photons is strongly suppressed despite nearly half the atoms being in
the excited state.  In the intermediate regime, $\Gamma_c < w < N
\Gamma_c$, the atoms emit light in a superradiant fashion and in the
strong pumping limit, $w \gg N \Gamma_c$, nearly all atoms are in the
excited state and they emit chaotic light like an ensemble of
thermally excited atoms.

\subsection{Monte-Carlo results for $g^{(2)}(0)$}

\begin{figure}
  \includegraphics{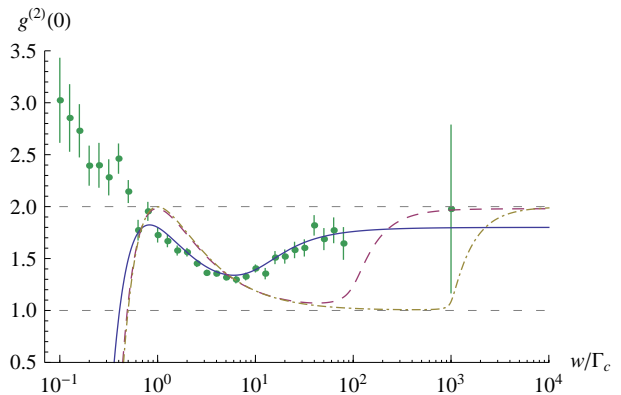}
  \caption{\label{fig:g2Zero} (Color online) Second order intensity
    correlation $g^{(2)}(0)$ as a function of the repump rate. The
    green symbols show the Monte-Carlo results including the
    statistical errors for $N=10$ atoms. The blue solid line shows the
    analytical result Eq. (\ref{eqn:analyticalG2}) for $N=10$ atoms,
    the purple dashed line is for $N=100$ atoms, and the yellow
    dash-dotted line is for $N=1000$ atoms. The gray dashed lines at
    $g^{(2)}(0)=2$ and $g^{(2)}(0)=1$ are for orientation.}
\end{figure}

First we consider $g^{(2)}(0)$ shown in Fig. \ref{fig:g2Zero} for
$N=10$ atoms.  The error bars in that figure are estimates of the
statistical uncertainty obtained by treating the histograms
$n_{i,j}(\Delta t)$ for different $i$ as independent of each other.

In the weak pumping limit the light exhibits strongly super-Poissonian
fluctuations indicating photon bunching.  This bunching effect can
easily be understood in the extreme limit $w/\Gamma_c\to
0$~\cite{Meiser:SteadyStateSuperradiance,Temnov:PhotonStatistics}.  In
that limit the atoms are optically pumped into collective dark
states~\cite{Benivegna:SubradiantStates,
  Nicolosi:SubradiantDickeBehaviour} $|J=1,M=-1\rangle$ and
$|J=0,M=0\rangle$~\footnote{For simplicity we only consider the case
  of an even total number of atoms here.  The argument applies with
  trivial modifications also to the case of an odd number of atoms.}.
From $\ket{J=0,M=0}$ the atoms can only be pumped to $\ket{J=1,M=1}$
from which they relax to $\ket{J=1,M=-1}$ by rapidly emitting a pair
of photons in a cascade within a time of order $\Gamma_c^{-1}$.

In the superradiant regime for intermediate pumping $g^{(2)}(0)$ is
reduced to a value about half way between 1 and 2.  It reaches a
minimum at the superradiant emission maximum, $w=N\Gamma_c/2$.  The
minimum value depends on $N$ with larger $N$ yielding smaller
$g^{(2)}(0)$.

In the limit of very strong repumping the atoms behave like a thermal
ensemble of atoms.  Therefore we have $g^{(2)}(0)=2(1-1/N)$ in that
limit in very good agreement with the Monte-Carlo simulation results.

\subsection{Semi-classical results for $g^{(2)}(0)$}

In applications, the number of atoms will typically be much larger than
1.  That makes it easier to achieve the collective strong coupling
regime where $N\mathcal{C} \gg 1$.  Furthermore, the collective nature
of the light emission is more apparent in that limit.  

Unfortunately, the Monte-Carlo simulations on which the above results
were based, cannot be easily implemented for the study of large atom
numbers.  The reason for this is that the size of the system Hilbert
space, $d = 2^N$, scales exponentially with the number of atoms.  In
this section, in order to bypass these difficulties, we exploit the
possibility for a semiclassical approximation that precisely derives
from the largeness of $N$.  The key idea is that in a macroscopic
ensemble of atoms the correlations between $n$ atoms can be expressed
in terms of the correlations between $n-1$ atoms plus an error term.
Ordinarily, the error terms become smaller as the ``cluster
size'' $n$ is increased.  The approximate treatment that we employ
here assumes that correlations between more than two atoms can be
completely expressed in terms of pair correlations and single atom
quantities.

The semi-classical calculation involves two non-trivial steps.  First
we have to find the correlations between the spins of different atoms.
Second we have to find an expression for $g^{(2)}(0)$ in terms of the
atomic correlations.

\subsubsection{Steady state solutions for pair correlations}

The symmetry of the expectation values with respect to particle
exchange greatly reduces the number of expectation values that have to
be considered.  We have for instance $\langle \hat \sigma_+^{(i)}\hat
\sigma_-^{(j)} \rangle = \langle \hat \sigma_+^{(1)}\hat
\sigma_-^{(2)}\rangle$ for all $i\neq j$.  Up to the level of pair
correlations, all observables that we are interested in can be
expressed in terms of $\langle \hat \sigma_z^{(1)}\rangle$, $\langle
\hat \sigma_+^{(1)}\hat \sigma_-^{(2)}\rangle$, and $\langle \hat
\sigma_z^{(1)}\hat \sigma_z^{(2)}\rangle$.

The equations of motion for these expectation values can be found from the
master equation (\ref{eqn:masterEqnJustAtoms}),
\begin{align}
  \frac{d\langle \hat
    \sigma_z^{(1)}\rangle}{dt}&= \label{eqn:sigmaz}
  -(w+\Gamma_c)\left(\langle \hat  \sigma_z^{(1)}\rangle-d_0\right)\\
  &\quad -2\Gamma_c(N-1)\langle \hat \sigma_+^{(1)}\hat
  \sigma_-^{(2)}\rangle\;,\nonumber
\end{align}
where $d_0=(w-\Gamma_c)/(w+\Gamma_c)$,
\begin{align}
  \frac{d\langle \hat\sigma_+^{(1)}\hat \sigma_-^{(2)}\rangle}{dt}&=
  -(w+\Gamma_c)\langle \hat
  \sigma_+^{(1)}\sigma_-^{(2)}\rangle\\
  &\quad+\frac{\Gamma_c}{2}\left(\langle\hat
    \sigma_z^{(1)}\hat\sigma_z^{(2)}\rangle+\langle \hat
    \sigma_z^{(1)}\rangle\right)\nonumber\\
  &\quad+\Gamma_c(N-2)\langle
  \hat\sigma_z^{(1)}
  \hat\sigma_+^{(2)}
  \hat\sigma_-^{(3)}
  \rangle\nonumber\;,
\end{align}
and
\begin{align}
  \frac{d\langle \hat \sigma_z^{(1)}\hat \sigma_z^{(2)}\rangle}{dt}&=
  \label{eqn:sigmaz1sigmaz2} -2(w+\Gamma_c)
  \left(
    \langle
    \hat \sigma_z^{(1)}\hat \sigma_z^{(2)}
    \rangle -
    d_0\langle \hat \sigma_z^{(1)}\rangle
  \right)\\
  &\quad +4\Gamma_c\Big(\langle \hat \sigma_+^{(1)}\hat
  \sigma_-^{(2)}\rangle\nonumber\\
&\qquad\qquad-(N-2)\langle \hat \sigma_z^{(1)}\hat \sigma_+^{(2)}\hat
  \sigma_-^{(3)}\rangle\Big)\;.\nonumber
\end{align}

\begin{figure}
\includegraphics[width=0.98\columnwidth]{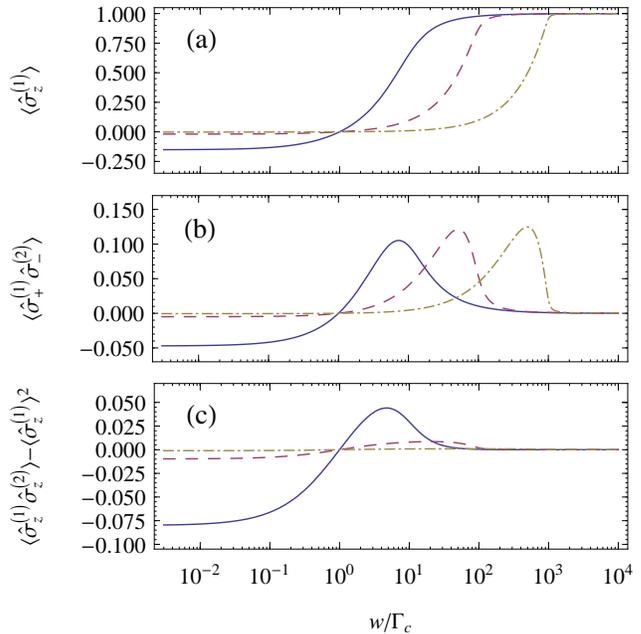}
\caption{\label{fig:inversionPlot}(Color online) (a) Inversion
  $\langle \hat \sigma_z^{(1)}\rangle$, (b) spin-spin correlation
  $\langle \hat\sigma_+^{(1)}\hat\sigma_-^{(2)}\rangle$, and (c)
  spin-spin correlation $\langle
  \hat\sigma_z^{(1)}\hat\sigma_z^{(2)}\rangle-
  \langle\hat\sigma_z^{(1)}\rangle^2$ as a function of pump strength
  for $N=10$ (blue solid line), $N=100$ (purple dashed line), and
  $N=1000$ atoms (yellow dash-dotted line).}
\end{figure}

We have checked that the third order expectation values $\langle \hat
\sigma_z^{(1)}\hat \sigma_+^{(2)}\hat \sigma_-^{(3)}\rangle$ can
be factorized according to
\[
\langle \hat
\sigma_z^{(1)}\hat \sigma_+^{(2)}\hat \sigma_-^{(3)}\rangle \approx
\langle \hat
\sigma_z^{(1)}\rangle\langle\hat \sigma_+^{(1)}\hat \sigma_-^{(2)}\rangle
\]
to a very good approximation by evaluating them in our Monte-Carlo
simulations and by approximating them in terms of lower order
cumulants in the adiabatic approximation.  By factorizing this way we
obtain a closed set of equations.  The steady state expectation values
are obtained by setting the time derivatives equal to zero.  The
resulting algebraic equations can be solved analytically for $\langle
\hat \sigma_z^{(1)}\rangle$, $\langle \hat \sigma_+^{(1)}\hat
\sigma_-^{(2)}\rangle$, and $\langle \hat \sigma_z^{(1)}\hat
\sigma_z^{(2)}\rangle$, leading to relatively complicated expressions
that we reproduce in the appendix for completeness.  Plots of the
steady state expectation values are given in
Fig. \ref{fig:inversionPlot} for different atom numbers.  For large
atom numbers the inversion $\langle \hat \sigma_z\rangle$ is
essentially zero below threshold for collective emission, increases
linearly with $w$ in the superradiant regime, and saturates with all
atoms in the excited state in the strong pumping regime.  In the
superradiant regime, the spin-spin correlation $\langle \hat
\sigma_+^{(1)}\sigma_-^{(2)}\rangle$ is approximately described by an
inverted parabola with zeros at the thresholds $w=\Gamma_c$ and
$w=N\Gamma_c$ and a peak value of $1/8$ at $w=N\Gamma_c/2$.  In the
strong pumping regime these correlations are destroyed by the
repumping.  The spin-spin correlations $\langle \hat
\sigma_z^{(1)}\hat \sigma_z^{(2)}\rangle$ approximately factorize as
$\langle \hat \sigma_z^{(1)}\hat \sigma_z^{(2)}\rangle\approx \langle
\hat \sigma_z^{(1)}\rangle^2$ in the limit of large atom numbers.
Note that the equations for $\langle \hat \sigma_z^{(1)}\rangle$ and
$\langle \hat \sigma_+^{(1)}\hat\sigma_-^{(2)}\rangle$ close if that
factorization is made and consequently much simpler approximate
expressions can be obtained for $w>\Gamma_c$ in the limit $N\to
\infty$ as pointed out in~\cite{Meiser:SteadyStateSuperradiance}.
However, since we would like to compare the semi-classical theory with
the Monte-Carlo results that were obtained for only relatively small
atom numbers, we have to use the more complicated expressions
discussed here.

\subsubsection{Expression of $g^{(2)}(0)$ in terms of atomic operators
  in the bad cavity limit}

In the bad cavity limit discussed in subsection
\ref{SubSec:AdiabaticElimination} the cavity field is slaved to
the atomic dipoles,
\begin{equation}
  \hat a \cong \frac{g}{i \kappa} \hat J_-\;.
\end{equation}
This means that we can calculate correlation functions of the field
if we know atomic correlation functions, for instance
\begin{align}
  g^{(2)}(0)
  &=\frac{\langle\hat J_+\hat J_+\hat J_-\hat J_-\rangle}{\langle \hat J_+\hat J_-\rangle^2}\notag\\
  &=\frac{\sum_{i,j,k,l=1}^N
    \langle \hat \sigma_+^{(i)}  \hat \sigma_+^{(j)}
    \hat \sigma_-^{(k)}  \hat \sigma_-^{(l)}\rangle}{
    \left(\sum_{i,j=1}^N \langle \hat \sigma_+^{(i)}
      \hat \sigma_-^{(j)}\rangle\right)^2}\;.
  \label{eqn:analyticalG2}
\end{align}
The atomic expectation values can be expressed in terms of the above expectation values,
\begin{equation}
  \sum_{i,j=1}^N\langle \hat \sigma_+^{(i)}\hat \sigma_-^{(j)}\rangle =
  N(\langle \hat \sigma_z^{(1)}\rangle+1)/2 +
  N(N-1)\langle\hat \sigma_+^{(1)}\hat \sigma_-^{(2)}\rangle\;,
\end{equation}
and
\begin{eqnarray}
  \langle \hat J_+\hat J_+\hat J_-\hat J_-\rangle& =&
  N(N-1)\\
  &&\times\Big(2(N-2)(\langle \hat \sigma_z^{(1)}\rangle+1)
  \langle \hat \sigma_+^{(1)}\hat \sigma_-^{(2)}\rangle\nonumber \\
  && \quad+(1+\langle \hat \sigma_z^{(1)}\hat \sigma_z^{(2)}\rangle+
  2\langle\hat\sigma_z^{(1)}\rangle^2)/2\nonumber\\
  && \quad+(N-2)(N-3)\langle \hat \sigma_+^{(1)}\hat \sigma_-^{(2)}\rangle^2\Big)\;.\nonumber
\end{eqnarray}
In order to arrive at this last result we have factorized expectation
values for four different atoms according to $\langle \hat
\sigma_+^{(1)}\hat\sigma_+^{(2)}\hat \sigma_-^{(3)}\hat
\sigma_-^{(4)}\rangle \approx \langle \hat
\sigma_+^{(1)}\hat\sigma_-^{(2)}\rangle^2$.  Expectation values in
which at least two indices are identical involve at most three
different atoms and, with the factorization discussed earlier, they
reduce to the known pair correlations and single atom expectation
values.

\subsubsection{Comparison with Monte-Carlo results}

The semi-classical results for $g^{(2)}(0)$ are also shown in
Fig. \ref{fig:g2Zero}.  The semi-classical curve agrees very well with
the Monte-Carlo results for $w>\Gamma_c$.  Below that threshold the
semi-classical expression yields unphysical values.  The disagreement
below threshold is not surprising because the atoms are in a very
highly correlated state in that regime and these correlations cannot
be captured by taking into account only pair-wise correlations.

The good agreement between semi-classical and Monte-Carlo results for
$w>\Gamma_c$ allows us to use the semi-classical expression to
extrapolate to very large atom numbers.  We find that for large atom
numbers the field exhibits nearly coherent counting statistics, {\it
  i.e.} $g^{(2)}(0)\approx 1$, in the superradiant regime,
$\Gamma_c\ll w<N\Gamma_c$, and it has the counting statistics of chaotic
light in the strong pumping regime.

\subsection{Monte-Carlo results for $g^{(2)}(\tau)$}

\begin{figure}
\includegraphics[width=8cm]{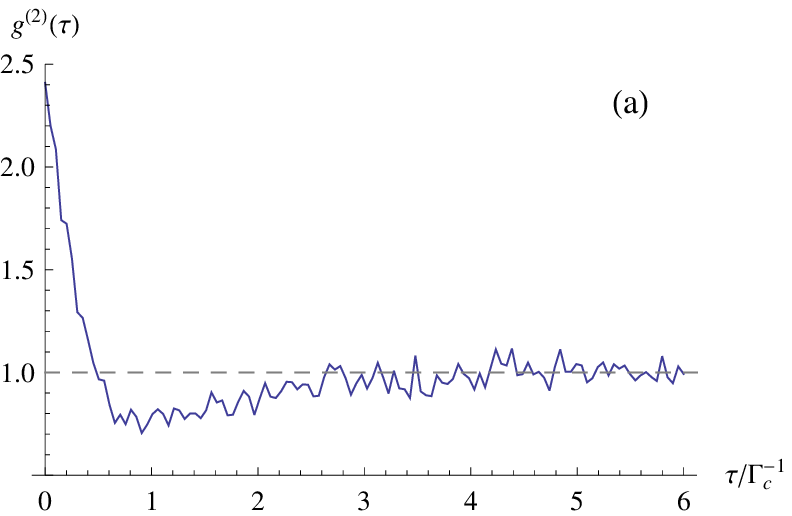}\\[0.2cm]
\includegraphics[width=8cm]{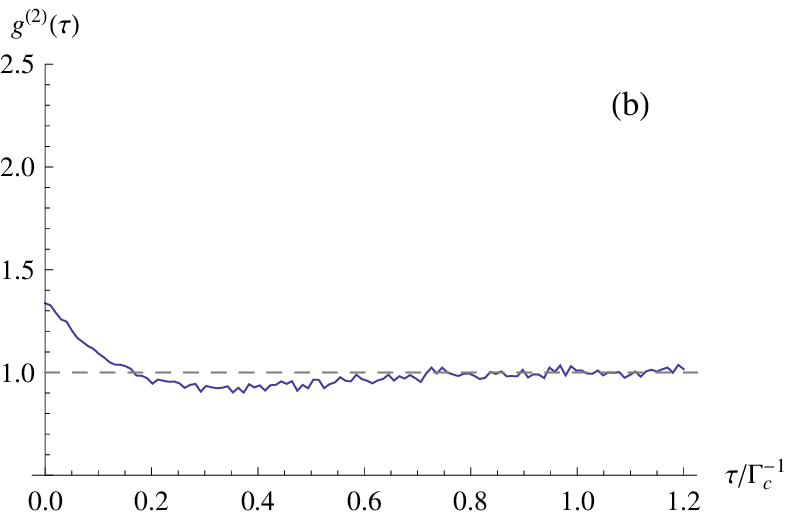}\\[0.2cm]
\includegraphics[width=8cm]{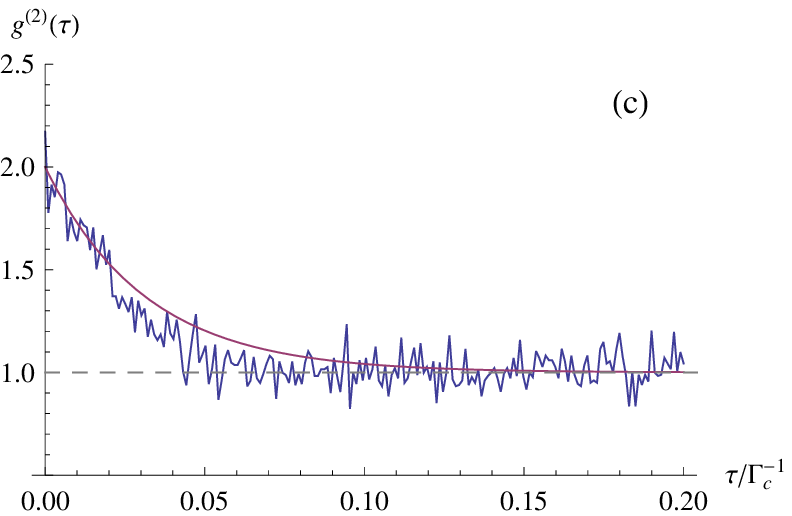}
\caption{\label{fig:G2OfTau} (Color online) Second order correlation
  $g^{(2)}(\tau)$ for $N=10$ atoms (a) for the subradiant regime for
  weak pumping $w=0.25\Gamma_c$, (b) for the superradiant regime with
  $w=5.0\Gamma_c$, and (c) for the strong pumping regime with
  $w=100\Gamma_c$. Note the different scales on the time axis for each
  figure.
  The purple line in (c) is the thermal light result.}
\end{figure}

The correlations between photons arriving at a photodetector
characterized by $g^{(2)}(0)$ only persist for a certain amount of
time.  To study the decay of these intensity correlations we show the
second order correlation for non-zero delay, $\tau \neq 0$, for 10
atoms in Fig. \ref{fig:G2OfTau}.

The strong bunching peak in the subradiant regime decays on a time
scale of order $\Gamma_c^{-1}$~\cite{Temnov:PhotonStatistics}.  After
that an anti-correlation dip develops for a period of order $\sim
w^{-1}$ because the repumping has to take the system out of the dark
states before another photon can be emitted.

In the superradiant regime the bunching is much weaker.  The bunching
peak also disappears on a much shorter time scale of order $\sim
1/(N\Gamma_c)$.  The subsequent anti-correlation is a much weaker
effect that disappears as $1/N$ in the limit of large atom numbers.
For times $\tau \gg w^{-1}, (N\Gamma_c)^{-1}$ the light intensities are
uncorrelated.  This is remarkable because the \emph{amplitude} of the
field is coherent for much longer times of order $\Gamma_c^{-1}$.
Intensity correlations decay on a time scale set by the collective
decay of the system while the decay of the amplitude correlations
occurs on the time scale set by the single particle decay.  From the
perspective of potential applications of this light source as an
ultra-stable local oscillator this means that the field can be
considered as coherent to a very good approximation.

As pointed out above the atoms behave like a thermal ensemble in the
strong pumping regime.  The two-time correlation function agrees well
with the result for thermal light emitted by a large number of
atoms~\cite{Loudon},
\begin{equation}
  g^{(2)}(t) \approx 1+|g^{(1)}(\tau)|^2 \approx
  1+e^{-\frac{2t}{2\pi/w}}\;.
  \label{thermalG2}
\end{equation}
The slight discrepancy between this formula and the numerical results
is likely due to the relatively small number of atoms considered while
Eq.(\ref{thermalG2}) is derived in the limit $N\to \infty$.

\section{Conclusion}
\label{sec:Conclusion}

The key result of this paper is that the light emitted in steady state
superradiance is second order coherent in the limit of large atom
numbers.  This result is significant because it establishes that the
coherence properties of the emitted light are closer to those of a
laser than those of light generated in ordinary pulsed superradiance.
In contrast, light generated in pulsed superradiance would have
super-Poissonian intensity fluctuations.  Such excess fluctuations
could adversely affect the utility of light sources based on steady
state superradiance as a stable frequency reference, one of the main
motivations for studying this system in the first place.  For such
applications it is crucial that the long coherence time and
collectively enhanced brightness demonstrated previously
\cite{Meiser:SrLaser} are paired with small intensity fluctuations.

In the subradiant regime the emitted light exhibits strong bunching
and super-Poissonian intensity fluctuations.  As has been pointed out
previously by Temnov and Woggon \cite{Temnov:PhotonStatistics} this
effect could be useful in identifying and analyzing the subradiant
regime in experiment.  The well understood thermal character of the
atomic ensemble in the strong pumping regime serves as a valuable
benchmark for the validity of our theoretical treatment.

In future research we plan to systematically investigate the
cross-over from steady state superradiance to a laser.  In the extreme
limits of this cross-over the system is dominated by purely atomic
collective enhancement on the one hand and by purely photonic
collective enhancement through stimulated emission on the other hand.
The intermediate regime between the two were both collective enhancement
due to stimulated emission and an atomic collective state are equally
important is very intriguing from a fundamental point of view.

\acknowledgments{We are grateful for stimulating discussions with
  J. K. Thompson, Jun Ye, and J. Cooper.  This work has been supported
  by NSF and ARO.}

\appendix

\section{Steady-state atomic pair correlations}
\label{sec:PairCorrelations}

In this appendix we summarize the analytical solution of the
semi-classical equations
Eqs. (\ref{eqn:sigmaz}-\ref{eqn:sigmaz1sigmaz2}).

In steady state, the atomic inversion is
\begin{equation}
  \langle \hat
  \sigma_z^{(1)}\rangle=
  d_0-\frac{2\Gamma_c(N-1)}{w+\Gamma_c}
  \langle \hat \sigma_+^{(1)}\hat  \sigma_-^{(2)}\rangle\;.
  \label{ssInversion}
\end{equation}
Inserting that into Eq. (\ref{eqn:sigmaz1sigmaz2}) we end up with a
linear equation for $\langle \hat \sigma_z^{(1)}\hat
\sigma_z^{(2)}\rangle$.  Solving that equation for $\langle \hat
\sigma_z^{(1)}\hat \sigma_z^{(2)}\rangle$ yields
\begin{align}
  \label{ssZZ}
  \langle \hat \sigma_z^{(1)}\hat \sigma_z^{(2)}\rangle &= 
d_0^2+\frac{2\Gamma_c}{(w+\Gamma_c)^2}
\Big(
(w+\Gamma_c)(1-d_0(2N-3))\notag\\
&\quad
+2(N-1)(N-2)\Gamma_c\langle \hat \sigma_+^{(1)}\hat\sigma_-^{(2)}\rangle\Big)
\end{align}
Inserting Eq. (\ref{ssInversion}) and Eq. (\ref{ssZZ}) into the
remaining equation for $\langle \hat \sigma_+^{(1)}\hat
\sigma_-^{(2)}\rangle$ leads to a quadratic equation.  One of the
solutions must be discarded because it violates $|\langle \hat
\sigma_+^{(1)}\hat \sigma_-^{(2)}\rangle_c| \leq 1$ for certain
repump rates and hence is unphysical.  The physically acceptable
solution is
\begin{widetext}
\begin{multline}
  \langle \hat\sigma_+^{(1)}\hat \sigma_-^{(2)}\rangle =
  -\frac{w+\Gamma_c}{4(N-1)(N-2)w\Gamma_c^2}\Big[
    w^2 + (2-(N-2)d_0)w\Gamma_c
+ (N-1)(1+d_0)\Gamma_c^2\\
-   \sqrt{4d_0(1+d_0)(N-1)(N-2)w\Gamma_c^3 +
      \left(w^2 + (2 - (N - 2)d_0) w \Gamma_c + (N-1)(1 + d_0)\Gamma_c^2\right)^2}
  \Big]\;.
\end{multline}
\end{widetext}
This solution can then be inserted in Eq.(\ref{ssInversion}) and
Eq. (\ref{ssZZ}) to find $\langle \hat \sigma_z^{(1)}\rangle_c$ and
$\langle \hat \sigma_z^{(1)}\hat\sigma_z^{(2)}\rangle_c$.

\bibliographystyle{apsrev}
\bibliography{mybibliographyEtAl}

\end{document}